\documentclass{ian}

\usepackage{dsfont}
\usepackage{array}
\usepackage{xcolor}
\usepackage{graphicx}

\definecolor{Complete Equation}{HTML}{4B0082}
\definecolor{Big Equation}{HTML}{DAA520}
\definecolor{Medium Equation}{HTML}{32CD32}
\definecolor{Simple Equation}{HTML}{4169E1}
\definecolor{Quantum Monte Carlo}{HTML}{DC143C}

\def\eqformat#1{{\bf\color{#1}#1}}

\begin{document}

\pagestyle{empty}

\hbox{}
\hfil{\bf\LARGE
Review of a Simplified Approach\par
\vskip10pt
\hfil to study the Bose gas at all densities
}
\vfill

\hfil{\bf\large Ian Jauslin}\par
\hfil{\it Department of Mathematics, Rutgers University}\par
\hfil{\tt\color{blue}\href{mailto:ian.jauslin@rutgers.edu}{ian.jauslin@rutgers.edu}}\par

\vfill

{\it
\hfill This paper is dedicated to Elliott, from whom I have learned so much;\par
\hfill with heartfelt appreciation and admiration, happy 90th birthday!
}
\vfill

\hfil {\bf Abstract}\par
\medskip
{\small
In this paper, we will review the results obtained thus far by Eric A. Carlen, Elliott H. Lieb and me on a Simplified Approach to the Bose gas.
The Simplified Approach yields a family of effective one-particle equations, which capture some non-trivial physical properties of the Bose gas at {\it both} low and high densities, and even some of the behavior at intermediate densities.
In particular, the Simplified Approach reproduces Bogolyubov's estimates for the ground state energy and condensate fraction at low density, as well as the mean-field estimate for the energy at high densities.
We will also discuss a phase that appears at intermediate densities with liquid-like properties.
The simplest of the effective equations in the Simplified Approach can be studied analytically, and we will review several results about it; the others are so far only amenable to numerical analysis, and we will discuss several numerical results.
We will start by reviewing some results and conjectures on the Bose gas, and then introduce the Simplified Approach and its derivation from the Bose gas.
We will then discuss the predictions of the Simplified Approach and compare these to results and conjectures about the Bose gas.
Finally, we will discuss a few open problems about the Simplified Approach.
}
\vfill

\tableofcontents

\vfill
\eject

\setcounter{page}1
\pagestyle{plain}

\section{Introduction}

\indent
The Bose gas is a fundamental model of quantum statistical mechanics.
It describes systems of many Bosons, from non-interacting photons in black body radiation to superfluid Helium IV.
Whereas in the case of photons the model is exactly solvable, helium atoms interact, which makes such a system both difficult to study, and gives it interesting properties.
In this paper, we will discuss a {\it Simplified Approach}, which was introduced by Lieb\-~\cite{Li63}, to study the Bose gas in its ground state {\it in the presence of interactions} in a tractable way.
We will discuss evidence that the Simplified Approach captures some of the physics of the interacting Bose gas, both in cases where the interaction between Bosons matters little, and in cases where it heavily affects the properties of the system.
More specifically, we will compare predictions of the Simplified Approach against known facts and widely accepted predictions for the Bose gas in the case of three observables: the {\it ground state energy}, the {\it condensate fraction} and the {\it two-point correlation function}.
\bigskip

\indent
The ground state energy of the repulsive Bose gas has been computed exactly in the regimes of asymptotically small and large densities\-~\cite{LY98,YY09,FS20,Li63}, which will allow us to make precise comparisons.
We will discuss analytical and numerical results that show that the predictions of the Simplified Approach for the energy are asymptotically correct at {\it both} low and high densities (at least in cases where the interaction between Bosons is of positive type, see below).
We will also see numerical evidence that some of the predictions of the Simplified Approach are extremely accurate at {\it all} densities.
\bigskip

\indent
While the energy is a valuable physical observable, the condensate fraction is arguably even more interesting.
In the early days of quantum mechanics, it had already been predicted\-~\cite{Bo24,Ei24} that Bosons would spontaneously form a phase called a {\it Bose-Einstein condensate} in which a positive fraction of particles are all in the same quantum state.
This was first demonstrated experimentally over 70 years after the prediction\-~\cite{AEe95,DMe95} in ultracold gases of rubidium and sodium.
The condensate fraction is the fraction of particles that are in this common quantum state.
From a mathematical point of view, Bose-Einstein condensation can be shown to occur spontaneously in systems of non-interacting Bosons\-~\cite{LSe05}, but there is, so far, no such proof for interacting Bosons at finite density in the continuum.
(Bose-Einstein condensation has been proved for Bosons on a lattice\-~\cite{KLS88} as well as in the Gross-Pitaevskii scaling regime\-~\cite{LS02,BBe18} which corresponds to zero-density.)
We will see that the Simplified Approach predicts Bose-Einstein condensation at low and high densities, and reproduces the widely accepted predictions of Bogolyubov theory.
\bigskip

\indent
The two-point correlation function is a measure of correlations between particles, and will be used to study the Bose gas in a range of densities where interactions play a preponderant role.
We will see that, at intermediate densities, the Simplified Approach predicts a liquid-like phase, whose existence we confirmed numerically for the Bose gas.
Furthermore, we will see that the two-point correlation function decays at all densities algebraically, as $|x|^{-4}$, which is compatible with the predictions of Bogolyubov theory.
\bigskip

\indent
In section\-~\ref{sec:Bose}, we will review known results and predictions about the Bose gas.
In section\-~\ref{sec:simplified}, we will introduce the Simplified Approach, and discuss its derivation from the Bose gas.
In section\-~\ref{sec:results}, we will review our results on the Simplified Approach, and provide links to the papers in which these are proved.
In section\-~\ref{sec:conclusion}, we discuss open problems.
In Appendix\-~\ref{app:simplesolv}, we discuss the software package {\tt simplesolv} that was written to carry out the numerical computations presented here, and give some details on the algorithms used.

\section{The Bose gas}\label{sec:Bose}
\indent
In this section, we will first introduce the formalism used to model the Bose gas, and review some theorems and widely accepted predictions about it.
\bigskip

\subsection{Definitions}

\indent
We consider a system of $N$ identical Bosons in a three-dimensional box of volume $V$.
We will take periodic boundary conditions, that is, the positions of the Bosons are in the three-dimensional torus of volume $V$: $\mathbb T^3:=\mathbb R^3/(V^{\frac13}\mathbb Z^3)$.
The wavefunction $\psi$ of the system is a symmetric function of $N$ positions: $\psi\in\mathcal H_N:=L_{2,\mathrm{symmetric}}((\mathbb T^{3})^N)$.
The Hamiltonian of the system is the self-adjoint operator on $\mathcal H_N$ defined by
\begin{equation}
  H_N=-\frac12\sum_{i=1}^N\Delta_i+\sum_{1\leqslant i<j\leqslant N}v(x_i-x_j)
\end{equation}
where $\Delta_i$ is the Laplacian with respect to the $i$-th position variable, and $v$ is the potential, which is a function in $L_1(\mathbb R^3)$ that is invariant under rotations, and satisfies
\begin{equation}
  v(x_i-x_j)\geqslant 0
  .
  \label{repulsive}
\end{equation}
The condition\-~(\ref{repulsive}) corresponds to the fact that the interaction $v$ is repulsive.
Note that we have taken the mass of each Boson to be $1$, whereas other references set the mass to $\frac12$.
\bigskip

\indent
We will focus solely on the ground state (that is, the system at zero-temperature), which we denote by $\psi_N$.
The ground state energy is $E_N$:
\begin{equation}
  H_N\psi_N=E_N\psi_N
  .
  \label{eigval}
\end{equation}
We take the thermodynamic limit, in which $N,V\to\infty$ in such a way that the density $\rho:=\frac NV$ is fixed, and define the ground state energy per particle:
\begin{equation}
  e_0:=\lim_{\displaystyle\mathop{\scriptstyle N,V\to\infty}_{\frac NV=\rho}}\frac{E_N}N
  .
  \label{gse}
\end{equation}
\bigskip

\indent
The condensate fraction is defined as the proportion of particles in the constant state $\varphi_{\mathrm c}:=V^{-\frac12}$.
Let $P_i$ be the projector acting on $\mathcal H_N$ onto the subspace in which the $i$-th particle is in the state $\varphi_{\mathrm c}$:
\begin{equation}
  P_i:=\underbrace{\mathds 1\otimes\cdots\otimes\mathds 1}_{i-1}\otimes\left|\varphi_{\mathrm c}\right>\left<\varphi_{\mathrm c}\right|\otimes\underbrace{\mathds 1\otimes\cdots\otimes\mathds 1}_{N-i-2}
  .
\end{equation}
In the papers in which the results presented below are proved, we have chosen to talk about the {\it un}condensed fraction $\eta$, rather than the condensate fraction, which is equal to $1-\eta$.
In order to avoid confusion, we will use the same convention here.
We define the uncondensed fraction:
\begin{equation}
  \eta=1-\lim_{\displaystyle\mathop{\scriptstyle N,V\to\infty}_{\frac NV=\rho}}\frac1N\sum_{i=1}^N\left<\psi_N\right|P_i\left|\psi_N\right>
  .
  \label{etadef}
\end{equation}
Note that $\eta$ can also be computed in terms of the ground state energy of a modified Hamiltonian: let
\begin{equation}
  K_N(\mu):=H_N-\mu\sum_{i=1}^NP_i
\end{equation}
and the associated ground state energy $F_N(N)$ satisfying $K_N(\mu)\Psi_N(\mu)=F_N(\mu)\Psi_N(\mu)$.
Since $\partial_\mu\Psi_N$ is orthogonal to $\Psi_N$,
\begin{equation}
  1-\frac1N\sum_{i=1}^N\left<\psi_N\right|P_i\left|\psi_N\right>
  =
  1+\frac1N\left.\partial_\mu\left<\Psi_N(\mu)\right|K_N(\mu)\left|\Psi_N(\mu)\right>\right|_{\mu=0}
  =
  \left.1+\partial_\mu\frac{F_N(\mu)}N\right|_{\mu=0}
  .
  \label{eta_energy}
\end{equation}
\bigskip

\indent
The two-point correlation function is defined with respect to the zero-temperature canonical Gibbs measure:
\begin{equation}
  C_2(y-y'):=\lim_{\displaystyle\mathop{\scriptstyle N,V\to\infty}_{\frac NV=\rho}}
  \sum_{i,j=1}^N\left<\psi_N\right|\delta(x_i-y)\delta(x_j-y')\left|\psi_N\right>
\end{equation}
where $\delta$ is the Dirac-delta function.
Similarly to the uncondensed fraction, $C_2$ can be computed by taking a functional derivative of the ground state energy with respect to the interaction potential.
To make this apparent, it is convenient to first use the translation invariance of the system to write $z\equiv y-y'$ and
\begin{equation}
  C_2(z)
  =
  \lim_{\displaystyle\mathop{\scriptstyle N,V\to\infty}_{\frac NV=\rho}}
  \frac1V\int dy'\ 
  \sum_{i,j=1}^N\left<\psi_N\right|\delta(x_i-z-y')\delta(x_j-y')\left|\psi_N\right>
\end{equation}
so, for $z\neq 0$,
\begin{equation}
  C_2(z)
  =
  \lim_{\displaystyle\mathop{\scriptstyle N,V\to\infty}_{\frac NV=\rho}}
  \frac2V\sum_{1\leqslant i<j\leqslant N}\left<\psi_N\right|\delta(x_i-x_j-z)\left|\psi_N\right>
  =
  \lim_{\displaystyle\mathop{\scriptstyle N,V\to\infty}_{\frac NV=\rho}}
  \frac2V\frac{\delta E_N}{\delta v(z)}
  =
  2\rho\frac{\delta e_0}{\delta v(z)}
  .
  \label{Cderiv}
\end{equation}

\subsection{Results and predictions}
\indent
The main difficulty in studying the repulsive Bose gas lies in the fact that Bosons interact.
One can get around this difficulty when the density is sufficiently low, since in that case particles interact little.
At very high densities, the Bose gas enters a {\it mean-field} regime\-~\cite{Se11}, in which the interactions can be replaced by an external field, and the problem reduces to a single-particle one which can readily be studied.
Because of this, we have both rigorous results and good approximation schemes in the low density and in the high density regimes.
\bigskip

\indent
To study the Bose gas at low densities, Bogolyubov\-~\cite{Bo47} introduced an approximation scheme that reduces the Hamiltonian to one that can be diagonalized explicitly\-~\cite{LSe05}, and thus leads to predictions about many observables, including the energy, condensate fraction, and two-point correlation function.
These were derived in a seminal paper by Lee, Huang and Yang\-~\cite{LHY57}, who found asymptotic expansions for the energy and condensate fraction, as well as the large distance behavior of the correlation function, among other results that will not be discussed here.
The expansion of the energy\-~(\ref{lhy}) has been the subject of much investigation, and has finally been proved under mild assumptions on the potential, which we state in the following Theorem.
\bigskip

\theoname{Theorem}{\cite{LY98,YY09,FS20,BCS21}}\label{theo:LHY}
  If $v\in L_3(\mathbb R^3)$ is non-negative, spherically symmetric, compactly supported, and has a bounded scattering length, then
  \begin{equation}
    e_0=2\pi\rho a\left(1+\frac{128}{15\sqrt\pi}\sqrt{\rho a^3}+o(\sqrt\rho)\right)
    \label{lhy}
  \end{equation}
  where $a$ is the scattering length of the potential\-~\cite[Appendix\-~C]{LSe05}.
\endtheo
\bigskip

The leading order $2\pi\rho a$ was derived by Lieb and Yngvason\-~\cite{LY98}, which lead to a revival in the study of the Lee-Huang-Yang formula, which had been introduced over forty years prior.
There were many improvements, leading first to an upper bound by Yau and Yin\-~\cite{YY09}, and later a lower bound by Fournais and Solovej\-~\cite{FS20}.
The assumptions on the potential in\-~\cite{YY09} were relaxed by Basti, Cenatiempo and Schlein\-~\cite{BCS21}.
The condition that $v\in L_3(\mathbb R^3)$ comes from the upper bound of\-~\cite{BCS21}; the lower bound of\-~\cite{FS20} only requires $v$ to be $L_1$.
This is a rather unusual situation, in which the assumptions for the upper bound are stronger than for the lower bound.
This discrepancy is even deeper than that: Fournais and Solovej have recently extended their lower bound to include hard-core interactions\-~\cite{FS21}, whereas the upper bound has not been proved to hold in that case (though it is expected to).
\bigskip

\indent
Bogolyubov theory predicts the following asymptotics for the uncondensed fraction at low densities.
\bigskip

\theoname{Conjecture}{\cite[(41)]{LHY57}}\label{conj:condensate}
  If $v\in L_1(\mathbb R^3)$ and $v\geqslant 0$, then as $\rho\to0$,
  \begin{equation}
    \eta\sim
    \frac{8\sqrt{\rho a^3}}{3\sqrt\pi}
  \end{equation}
  where $a$ is the scattering length of the potential\-~\cite[Appendix\-~C]{LSe05}.
\endtheo
\bigskip

In particular, this implies that, as $\rho\to0$, $\eta\to0$, which means that there is full condensation at zero-density.
As was mentioned above, there is, as of this writing, no proof that Bose-Einstein condensation occurs at any finite density, so this conjecture has no proof.
Recent progress has been made in the proof of complete condensation in the Gross-Pitaevskii regime\-~\cite{LS02,BBe18}, in which the interaction scales with the potential: the scattering length decreases with $N$, which corresponds to an ultra-dilute limit.
\bigskip

\indent
Bogolyubov theory further predicts that the two-point correlation function decays as $|x|^{-4}$.
\bigskip

\theoname{Conjecture}{\cite[(48)]{LHY57}}\label{conj:2pt}
  If $v\in L_1(\mathbb R^3)$ and $v\geqslant 0$, then as $\sqrt{\rho a}|x|\to\infty$,
  \begin{equation}
    \frac1{\rho^2}C_2(x)-1\sim\frac{16\rho a^3}{\pi^3(\sqrt{\rho a}|x|)^4}
  \end{equation}
  where $a$ is the scattering length of the potential\-~\cite[Appendix\-~C]{LSe05}.
\endtheo
\bigskip

\indent
At high densities, the system approaches a {\it mean-field} regime, and can readily be studied.
The asymptotics for the energy are significantly simpler than for low density, and were proved by Lieb in 1963.
\bigskip

\theoname{Theorem}{\cite[Appendix]{Li63}}\label{theo:Hartree}
  If $v\in L_1(\mathbb R^3)$, $v\geqslant 0$, and its Fourier transform $\hat v$ is non-negative ($v$ is then said to be of {\it positive type}), then, as $\rho\to\infty$,
  \nopagebreakaftereq
  \begin{equation}
    e_0\sim\frac\rho2\int dx\ v(x)
    .
    \label{Bose_high_density}
  \end{equation}
\endtheo
\restorepagebreakaftereq

The assumption that $\hat v\geqslant 0$ is necessary as one can find counterexamples in which $e_0$ grows slower than $\rho$ at high densities for potentials that are not of positive type.
\bigskip

\indent
At high densities, since the system approaches a mean-field regime, it is expected that it should be completely condensed.
\bigskip

\theo{Conjecture}
  If $v\in L_1(\mathbb R^3)$, then, as $\rho\to\infty$,
  \nopagebreakaftereq
  \begin{equation}
    \eta\to1
    .
  \end{equation}
\endtheo
\restorepagebreakaftereq

As in the low density regime, this conjecture has, as of this writing, not been proved.

\section{The Simplified Approach}\label{sec:simplified}
\indent
Let us now turn to the derivation of the Simplified Approach, following\-~\cite{Li63}.
This derivation will involve a significant approximation, which has not been justified rigorously as of this writing.
In this paper, we will not discuss how the approximation can be justified, rather the aim here is to study the Simplified Approach and use it as a tool to compute physical observables for the repulsive Bose gas.
Justifying this approximation will be the goal of future work.
\bigskip

\subsection{Derivation of the Simplified Approach from the Bose gas}
\indent
We start with the eigenvalue equation\-~(\ref{eigval}), and integrate both sides of the equation (which is formally identical to taking a scalar product with the constant function, which has non-trivial overlap with the condensate wavefunction), and find (using the symmetry under exchange of particles)
\begin{equation}
  \frac{E_N}N=\frac{(N-1)}2\int dx_1dx_2\ v(x_1-x_2)\frac{\int dx_3\cdots dx_N\ \psi_N(x_1,\cdots,x_N)}{\int dy_1\cdots dy_N\ \psi_N(y_1,\cdots,y_N)}
  \label{inteig}
\end{equation}
(note that the kinetic term vanishes, since it is the integral of an exact derivative).
The crucial observation is that, since $\psi_N$ is the ground state of the Hamiltonian, it is non-negative (this can be verified by checking that $|\psi|$ has the same energy as $\psi$, and then using the Perron-Frobenius theorem to prove the uniqueness of the ground state), and so $\psi/\int\psi$ can be interpreted as a probability distribution.
In that language, $\int dx_3\cdots dx_N\psi/\int dy_1\cdots dy_N\psi$ is the two-point correlation function of $\psi/\int\psi$.
When defining these correlation functions, we actually normalize the integrals, which will allow us to keep track of volume factors more readily, and define
\begin{equation}
  g^{(n)}_N(x_1,\cdots,x_n):=\frac{\int\frac{dx_{n+1}}V\cdots\frac{dx_N}V\ \psi_N(x_1,\cdots,x_N)}{\int\frac{dy_1}V\cdots\frac{dy_N}V\ \psi_N(y_1,\cdots,y_N)}
  \equiv
  \frac{V^n\int dx_{n+1}\cdots dx_N\ \psi_N(x_1,\cdots,x_N)}{\int dy_1\cdots dy_N\ \psi_N(y_1,\cdots,y_N)}
  \label{g}
\end{equation}
in terms of which\-~(\ref{inteig}) becomes, using the translation invariance of the system to eliminate the integral over $x_2$,
\begin{equation}
  \frac{E_N}N=\frac{N-1}{2V}\int dx\ v(x)g_N^{(2)}(x,0)
  .
\end{equation}
Taking the thermodynamic limit, we find
\begin{equation}
  e_0=\frac\rho2\int dx\ v(x)g^{(2)}(x)
  \label{energy_exact}
\end{equation}
where $g^{(2)}(x):=\lim_{N,V\to\infty} g_N^{(2)}(x,0)$.
\bigskip

\indent
We have thus reduced the question of computing $e_0$ to that of computing $g^{(2)}(x)$.
To do so, we proceed in a similar fashion as before by integrating the eigenvalue equation\-~(\ref{eigval}), but this time, we integrate over all variables but the first two, and find
\begin{equation}
  \begin{array}{>\displaystyle l}
    -\frac12(\Delta_x+\Delta_y) g_N^{(2)}(x,y)
    +\frac{N-2}V\int dz\ (v(x-z)+v(y-z))g_N^{(3)}(x,y,z)
    +\\[0.5cm]\indent\hfill+
    v(x-y)g_2(x,y)
    +\frac{(N-2)(N-3)}{2V^2}\int dzdt\ v(z-t)g_N^{(4)}(x,y,z,t)
    =E_Ng_N^{(2)}(x,y)
  \end{array}
  \label{eqg2}
\end{equation}
where $g^{(3)}_N$ and $g^{(4)}_N$ are defined in\-~(\ref{g}).
Thus, to compute $g^{(2)}$ in this way, we need to compute $g^{(3)}$ and $g^{(4)}$. One can iterate this procedure and define a {\it hierarchy} of equations to compute all the $g^{(n)}$, but this set of equations will be very difficult, if not impossible, to solve.
Note that, by their definition\-~(\ref{g}), the $g$'s are not independent: for all $m<n$,
\begin{equation}
  \int \frac{dx_{m+1}}V\cdots \frac{dx_n}V\ g_N^{(n)}(x_1,\cdots,x_n)
  =
  g_N^{(m)}(x_1,\cdots x_m)
  .
  \label{inducg}
\end{equation}
We will proceed by making an approximation to express $g^{(3)}$ and $g^{(4)}$ in terms of $g^{(2)}$, that preserves some of the equalities in\-~(\ref{inducg}).
\bigskip

\theo{Approximation}\label{approx:main}
  There exist two functions $w,h:\mathbb R^3\to\mathbb R$ such that
  \begin{equation}
    g_N^{(3)}(x,y,z)=(1-w(x-y))(1-w(x-z))(1-w(y-z))
    ,\hskip10pt
    g_N^{(4)}(x_1,x_2,x_3,x_4)=\prod_{1\leqslant i<j\leqslant 4}(1-h(x_i-x_j))
    \label{factorization}
  \end{equation}
  in which $w,h$ are chosen so that
  \begin{equation}
    \int \frac{dz}V\ g_N^{(3)}(x,y,z)
    =g_N^{(2)}(x,y)
    ,\quad
    \int \frac{dzdt}{V^2}\ g_N^{(4)}(x,y,z,t)
    =g_N^{(2)}(x,y)
    \label{cd34}
  \end{equation}
  and $\int dx\ |1-w(x)|$ and $\int dx\ |1-h(x)|$ are bounded uniformly in $V$.
\endtheo
\bigskip

Unfortunately, (\ref{cd34}) does not ensure that\-~(\ref{inducg}) is satisfied, since $\int \frac{dt}V\ g^{(4)}(x,y,z,t)$ will not, in general, be equal to $g^{(3)}(x,y,z)$.
One can then prove that\-~(\ref{cd34}) imposes an explicit expression of $g_N^{(3)}$ and $g_N^{(4)}$ in terms of $g_N^{(2)}$.
\bigskip

\theoname{Lemma}{\cite[(3.25),(3.28),(3.15)]{Li63}}
  Approximation\-~\ref{approx:main} implies that, provided $\int dx\ |1-g_N^{(2)}(x,0)|$ is bounded independently of $V$,
  \begin{equation}
    w(x-y)=1-g_N^{(2)}(x,y)+O(V^{-1})
  \end{equation}
  and
  \begin{equation}
    h(x-y)=1-g_N^{(2)}(x,y)+\frac2Vg_N^{(2)}(x,y)\int dz\ (1-g_N^{(2)}(x,z))(1-g_N^{(2)}(y,z))+O(V^{-2})
    .
    \label{h}
  \end{equation}
\endtheo
\bigskip

Note that it is important to keep the term of order $V^{-1}$ in\-~(\ref{h}), since $g_N^{(4)}$ appears in\-~(\ref{eqg2}) in a term that is of order $N$.
Thus, the approximation becomes
\begin{equation}
  g_N^{(3)}(x,y,z)\approx
  g_N^{(2)}(x,y)
  g_N^{(2)}(x,z)
  g_N^{(2)}(y,z)
  +O(V^{-1})
\end{equation}
and
\begin{equation}
  g_N^{(4)}(x_1,x_2,x_3,x_4)\approx
  \prod_{1\leqslant i<j\leqslant 4}g_N^{(2)}(x_i-x_j)+O(V^{-1})
  .
\end{equation}
Plugging these (with the $V^{-1}$ correction in\-~(\ref{h})) into\-~(\ref{eqg2}) and taking the thermodynamic limit, we find the ``{\it Complete Equation}'' of the Simplified Approach.
Before defining this equation, we first define
\begin{equation}
  u(x):=\lim_{\displaystyle\mathop{\scriptstyle N,V\to\infty}_{\frac NV=\rho}}(1-g_N^{(2)}(x,0))
  \label{udef}
\end{equation}
and recall the definition of the convolution operator
\begin{equation}
  f\ast g(x):=\int dy\ f(x-y)g(y)
  .
\end{equation}
\bigskip

\theoname{Definition}{Complete Equation of the Simplified Approach}\label{def:compleq}
  \begin{equation}
    -\Delta u(x)
    =
    (1-u(x))\left(v(x)-2\rho K(x)+\rho^2 L(x)\right)
    \label{compleq}
  \end{equation}
  \begin{equation}
    K:=
    u\ast S
    ,\quad
    S(y):=(1-u(y))v(y)
  \end{equation}
  \nopagebreakaftereq
  \begin{equation}
    L:=
    u\ast u\ast S
    -2u\ast(u(u\ast S))
    +\frac12
    \int dydz\ u(y)u(z-x)u(z)u(y-x)S(z-y)
    .
  \label{compleqL}
  \end{equation}
\endtheo
\restorepagebreakaftereq
\bigskip

\indent
Proceeding in this way, we have reduced the problem of the computation of the ground state energy of the Bose gas, by making Approximation\-~\ref{approx:main}, to solving a non-linear, non-local (because of the non-local nature of the convolution operator $\ast$) partial differential equation, that only involves functions of $\mathbb R^3$, that is we have reduced the problem to a single particle one.
This type of reduction is done in many other contexts in physics, such as the Boltzmann equation, the Gross-Pitaevskii equation, the Hartree equation, the Thomas-Fermi equation, etc...
However, one feature that makes the Simplified Approach stand out from these other effective equations is that, whenever an effective equation can be proved to be exact, it is usually in the limit of a parameter being very small or very large (for instance the Gross-Pitaevskii equation holds for very small densities\-~\cite{LS02}, the Hartree equation holds for very large densities\-~\cite{Se11}), whereas, as we will see in the next section, the Simplified Approach is exact in {\it both} the low density and the high density regime.
\bigskip

\indent
This can be anticipated from the nature of Approximation\-~\ref{approx:main}.
The factorization in\-~(\ref{factorization}), remembering that $g_N^{(n)}$ is an $n$-point correlation function of a classical probability, is an {\it independence} condition (in classical statistical mechanics, this is called {\it clustering}).
The fact that particles at low density should be approximately independent is reasonable, since interactions are few and far between.
At high density, the system should approach a mean-field regime, in which particles are effectively independent.
However, making these arguments rigorous is still an open problem.
In addition, as we will see in the next section, this approximation can also be quantitatively very good, even at intermediate densities, where the approximation would be quite difficult to justify.

\subsection{The Equations of the Simplified Approach}
\indent
Before discussing the prediction of the Simplified Approach, we must first discuss the various equations that it consists of.
So far, we have seen the {\it Complete Equation} in Definition\-~\ref{def:compleq}, but this equation is rather difficult to deal with, both analytically and numerically.
From a numerical point of view, it is costly to compute nested integrals: the method used to carry out the numerics is a Gaussian quadrature, in which each integral is replaced by a sum of $N$ terms, and taking an integral in $n$ variables requires $N^n$ points.
In practice, a quintuple integral is near the upper limit of what is computable quickly on ordinary hardware (as of this writing).
Several tricks are at our disposal to reduce the number of variables: in the case of a convolution, the integral is three dimensional, but by changing to {\it two-center bipolar coordinates}, we can reduce it to a double integral.
In addition, the equation in Definition\-~\ref{def:compleq} has fewer integrations in Fourier space, in which convolutions become products and vice versa.
These allow us to reduce the computation to a sextuple integral (for details, see the documentation of the {\tt simplesolv} software package\-~\cite{ss}), which is still rather heavy to compute.
However, the sextuple integral appears only in the last term in\-~(\ref{compleqL}), and if this term is dropped, we are left with only double integrals.
We thus define an equation obtained from the Complete Equation in which we drop the last term in\-~(\ref{compleqL}), which we will call the ``{\it Big Equation}''.
\bigskip

\theoname{Definition}{Big Equation of the Simplified Approach}
  \begin{equation}
    -\Delta u(x)
    =
    (1-u(x))\left(v(x)-2\rho K_{\mathrm{bigeq}}(x)+\rho^2 L_{\mathrm{bigeq}}(x)\right)
    \label{bigeq}
  \end{equation}
  \begin{equation}
    K_{\mathrm{bigeq}}:=
    u\ast S
    ,\quad
    S(y):=(1-u(y))v(y)
    \label{bigeqK}
  \end{equation}
  \nopagebreakaftereq
  \begin{equation}
    L_{\mathrm{bigeq}}:=
    u\ast u\ast S
    -2u\ast(u(u\ast S))
    .
  \label{bigeqL}
  \end{equation}
\endtheo
\bigskip

To evaluate the precision of this approximation, we have computed observables for the Complete Equation (the sextuple integral can be computed in under 24 hours provided the order $N$ is chosen to be small enough), and found very good agreement.
\bigskip

\indent
Whereas the Big Equation is (somewhat) easily computable numerically (see the documentation of the {\tt simplesolv} software package\-~\cite{ss}), it remains rather difficult to study analytically.
To simplify the equation, we will make two approximations.
First, we assume that $u(x)\ll 1$, which is correct in the limit $|x|\to\infty$.
This allows us to approximate
\begin{equation}
  (1-u)(-2\rho K+\rho^2L)\approx-2\rho K+\rho^2L
  ,\quad
  L\approx u\ast u\ast S
  .
  \label{simpleq_approx1}
\end{equation}
Second, we will replace $S$ by a Dirac delta function (while preserving its integral):
\begin{equation}
  S(x)\approx\delta(x)\int dx\ S(x)\equiv\frac{2\tilde e}\rho\delta(x)
  ,\quad
  \tilde e:=\frac\rho2\int dx\ S(x)\equiv\frac\rho2\int dx\ (1-u(x))v(x)
  .
\end{equation}
The idea behind this approximation is that, when $|x|$ is large, $S$ can be concentrated at the origin.
Using this second approximation, we get
\begin{equation}
  K\approx\frac{2\tilde e}\rho u
  ,\quad
  L\approx\frac{2\tilde e}\rho u\ast u
  \label{simpleq_approx2}
\end{equation}
which leads us to define the ``{\it Simple Equation}'' as follows.
\bigskip

\theoname{Definition}{Simple Equation of the Simplified Approach}
  \begin{equation}
    -\Delta u(x)=(1-u(x))v(x)-4\tilde e u(x)+2\tilde e\rho u\ast u(x)
    \label{simpleq}
  \end{equation}
  \nopagebreakaftereq
  \begin{equation}
    \tilde e:=\frac\rho2\int dx\ (1-u(x))v(x)
    .
  \end{equation}
\endtheo
\restorepagebreakaftereq
\bigskip

\indent
Finally, we define an intermediate equation, which is simpler than the Big Equation to compute, but has fewer approximations than the Simple Equation.
As we will see below, this ``{\it Medium Equation}'' agrees quantitatively rather well with the Big Equation and the Bose gas.
To define the Medium Equation, we make the approximation in\-~(\ref{simpleq_approx1}), but not\-~(\ref{simpleq_approx2}).
\bigskip

\theoname{Definition}{Medium Equation of the Simplified Approach}
  \begin{equation}
    -\Delta u(x)=(1-u(x))v(x)-2\rho u\ast S(x)+\rho^2u\ast u\ast S(x)
  \end{equation}
  \nopagebreakaftereq
  \begin{equation}
    S(x):=(1-u(x))v(x)
    .
  \end{equation}
\endtheo
\restorepagebreakaftereq

\section{Predictions of the Simplified Approach}\label{sec:results}
\indent
In this section, we state the results and predictions we obtained for the Simplified Approach.
We start with a theorem on the existence of solutions of the Simple Equation, and then move on to discuss predictions for the energy, condensate fraction, and two-point correlation function.
\bigskip

\subsection{Existence and Uniqueness}\label{sec:uniqueness}
\indent
In order to state any theorem about the solutions of the equations of the Simplified Approach, we first need to prove that these exist.
So far, we only have such a result for the Simple Equation, which holds in arbitrary dimension $d$.
\bigskip

\theoname{Theorem}{\cite[Theorems 1.1, 1.3]{CJL20}}\label{theo:existence}
  If $v\in L_1(\mathbb R^d)\cap L_p(\mathbb R^d)$ for $p>\max\{\frac d2,1\}$ and $v\geqslant 0$, then\-~(\ref{simpleq}) has an integrable solution $u$ satisfying $0\leqslant u(x)\leqslant 1$.
\endtheo
\bigskip

\indent
The issue of the uniqueness of the solution is a bit more subtle, and to state it, let us first discuss the main points of the proof of this theorem.
Equation\-~(\ref{simpleq}) contains two non-linear terms: $2\tilde e\rho u\ast u$ and $-4\tilde eu$, since $\tilde e$ depends on $u$.
However, we can simplify this situation by changing our point of view on this equation.
As is, the equation takes $\rho$ as a parameter, and returns both $\tilde e(\rho)$ and $u(x)$.
But if we fix $\tilde e$ as a parameter, and view\-~(\ref{simpleq}) as an equation that will return $\rho(\tilde e)$ and $u(x)$, then $-4\tilde e u(x)$ becomes a linear term (and conveniently opens a gap in the Laplacian operator).
Proceeding in this way, we can prove that the solution $(\rho,u(x))$ exists, {\it and is unique}\-~\cite[Theorem\-~1.3, Section\-~3]{CJL20}.
To prove Theorem\-~\ref{theo:existence}, we are left with proving that $\tilde e\mapsto\rho(\tilde e)$ can be inverted locally.
This is the case because $\rho(\tilde e)$ is continuous, and goes from $0$ to $\infty$\-~\cite[Theorem\-~1.3]{CJL20}.
However, proving that $\rho(\tilde e)$ is monotone is still open (though we expect it to be true), so we do not have a strong uniqueness statement, as it is {\it a priori} possible that a given value of $\rho$ yields several $\tilde e$'s.
On the other hand, the theorems stated in the following subsections hold for {\it any} solution of\-~(\ref{simpleq}), so if there were several solutions, they would share the properties we will state below.
\bigskip

\indent
In three dimensions, for small and large enough values of $\tilde e$, we have proved the monotonicity of $\rho(\tilde e)$.
\bigskip

\theoname{Theorem}{\cite[Theorem 1.3]{CJL21}}
  If $(1+|x|^4)v\in L_1(\mathbb R^3)\cap L_2(\mathbb R^3)$, $v\geqslant 0$, and if
  \begin{equation}
    \tilde e<\frac{\sqrt 2\pi^3}{\|v\|_1^2}
    \quad\mathrm{or}\quad
    \tilde e>\frac{8\|v\|_2^4}{\pi^4}
  \end{equation}
  then $\tilde e\mapsto\rho(\tilde e)$ is monotone.
\endtheo

The proof of this theorem relies on the study of the operator
\begin{equation}
  \mathfrak K_{\tilde e}:=(-\Delta+v+4\tilde e(1-\rho u\ast))^{-1}
  \label{frakK}
\end{equation}
which comes out naturally when differentiating $u$ with respect to $\tilde e$ (which is a natural thing to do to prove monotonicity).
In particular, the proof relies on an estimate of $\|\mathfrak K_{\tilde e}\psi\|_{\frac43}$\-~\cite[Section\-~1.2]{CJL21} for $\psi$'s that integrate to zero.
The Hardy-Littlewood-Sobolev (HLS) inequality would allow us to bound $\|\mathfrak K_{\tilde e}\psi\|_{\frac32+\epsilon}$, which is insufficient, but HLS does not make use of the fact that $\int\psi=0$.
We proved a variant of the HLS inequality for functions that integrate to zero\-~\cite[Theorem 1.13]{CJL21}, which yields a usable bound.

\subsection{Properties of the solution of the Simple Equation}
\indent
Let us now review a few interesting properties of the solution $u$ of the Simple Equation.
We start with properties that hold in all dimensions.
\bigskip

\theo{Lemma}
  Under the assumptions of Theorem\-~\ref{theo:existence}, in any dimension $d\geqslant 1$, if $u$ is an integrable solution of the Simple Equation, then
  \begin{itemize}
    \item $u(x)\leqslant 1$ if and only if $u(x)\geqslant 0$ \cite[Theorems 1.1, 1.2]{CJL20}.
    \item $\int dx\ u(x)=\frac1\rho$ \cite[(1.7)]{CJL20}.
    \item If $u(x)\geqslant 0$, then $u(x)\geqslant (-\Delta+4\tilde e+v)^{-1}v$ and $\frac\rho4\|v\|_1\leqslant\tilde e\leqslant\frac\rho2\|v\|_1$ \cite[(1.18),(1.22)]{CJL20}.
  \end{itemize}
\endtheo
\bigskip

\indent
From now on, and until the end of this paper, we will focus on three dimensions.
Let us start with the more significant result, which concerns the large $|x|$ behavior of $u$.
\bigskip

\theoname{Theorem}{\cite[Theorem\-~1.2]{CJL21}}\label{theo:decay}
  In the case $d=3$, if $u$ is a non-negative, integrable solution of the Simple Equation, $v\geqslant 0$, and $(1+|x|^4)v\in L_1(\mathbb R^3)\cap L_2(\mathbb R^3)$, then
  \begin{equation}
    \rho u(x)=\frac{\sqrt{2+\beta}}{2\pi^2\sqrt{\tilde e}}\frac1{|x|^4}+R(x)
  \end{equation}
  where $\beta:=\rho\int dx\ |x|^2v(x)(1-u(x))$ and $|x|^4R\in L_2(\mathbb R^3)\cap L_\infty(\mathbb R^3)$ uniformly in $\tilde e$ on all compact sets.
  Furthermore, for any $\rho_0>0$, if $\rho<\rho_0$
  \begin{equation}
    u(x)\leqslant \frac C{\rho\tilde e^{\frac12}|x|^4}
  \end{equation}
  for some $C$ that only depends on $\rho_0$.
\endtheo
\bigskip

Up to the error term $R$, this shows that $u$ decays as $|x|^{-4}$.
It is quite natural that $u$ should decay as a power law, since, at large $|x|$, $u$ must decay at the same rate as $u\ast u$ (see\-~(\ref{simpleq})), which is the case for power laws (but not for, say, an exponential).
One can also see that the power $4$ is to be expected: defining $f:=2\tilde e\rho(-\Delta+4\tilde e)^{-1}u$, one easily checks\-~\cite[(2.5)]{CJL20} that
\begin{equation}
  f\geqslant f\ast f
  ,\quad
  \int dx\ f(x)=\frac12.
  \label{fconv}
\end{equation}
Furthermore, one can prove that\-~\cite[Theorem\-~2]{CJe21} any $f$ satisfying\-~(\ref{fconv}) must also satisfy $\int dx\ |x|f(x)=\infty$, which suggests that $f$ decays like $|x|^{-4}$ (though it does not prove it: $\int dx\ |x|f(x)=\infty$ alone does not mean that $f\sim|x|^{-4}$), and consequently that $u$ does as well.
The proof of Theorem\-~\ref{theo:decay} follows a different line: we work in Fourier space, and study the small momentum behavior of $u$\-~\cite[Section\-~2]{CJL21}.

\bigskip

\indent
Before concluding this subsection, let us state some estimates on $u$ which have been used to prove some of the theorems mentioned in this paper.
\bigskip

\theo{Lemma}
  Under the assumptions of Theorem\-~\ref{theo:existence}, in the case $d=3$, if $u$ is a non-negative, integrable solution of the Simple Equation, then
  \begin{itemize}
    \item
    For $1\leqslant p<3$,
    \begin{equation}
      \|u\|_p\leqslant C_p\tilde e^{\frac{p-3}{2p}}\|v\|_1
    \end{equation}
    where $C_p$ only depends on $p$\-~\cite[Lemma\-~1.1]{CJL21}.

    \item
    For large $\rho$, we can improve the bound\-~\cite[Lemma\-~1.1]{CJL21}
    \begin{equation}
      \|u\|_2\leqslant\frac1{2\tilde e}\|v\|_2
      .
    \end{equation}

    \item
    We have
    \begin{equation}
      \|\partial_{\tilde e}u\|_2\leqslant \frac C{\rho\tilde e^{\frac14}}
    \end{equation}
    for some $C$ independent of $\tilde e$ (which may depend on $v$)\-~\cite[Lemma\-~1.11]{CJL21}.
  \end{itemize}
\endtheo

\subsection{Ground state energy}
\indent
The Simplified Approach provides a natural prediction for the ground state energy per-particle (see\-~(\ref{energy_exact}) and\-~(\ref{udef})):
\begin{equation}
  \tilde e=\frac\rho2\int dx\ (1-u(x))v(x)
  .
\end{equation}
We use the tilde in the notation to indicate that this is a prediction of the Simplified Approach, as it distinguishes it form the exact ground state energy of the Bose gas $e_0$.
\bigskip

\indent
As was explained in Section\-~\ref{sec:Bose}, the ground state energy of the Bose gas has been computed at low and high densities, see Theorems\-~\ref{theo:LHY} and\-~\ref{theo:Hartree}.
We have proved that both of these asymptotic expansions hold for the prediction of the Simple Equation.
\bigskip

\theoname{Theorem}{\cite[Theorem 1.4]{CJL20}}\label{theo:energy}
  For the Simple Equation, under the assumptions of Theorem\-~\ref{theo:existence} for $d=3$, as $\rho\to0$,
  \begin{equation}
    \tilde e=2\pi\rho a\left(1+\frac{128}{15\sqrt\pi}\sqrt{\rho a^3}+o(\sqrt\rho)\right)
  \end{equation}
  where $a$ is the scattering length of the potential\-~\cite[Appendix C]{LSe05}, and as $\rho\to\infty$
  \nopagebreakaftereq
  \begin{equation}
    \tilde e\sim_{\rho\to\infty}\frac\rho2\int dx\ v(x)
    .
    \label{high_density}
  \end{equation}
\endtheo
\restorepagebreakaftereq
\bigskip

It is rather striking that the Simple Equation agrees with the Bose gas {\it both} at high and low densities.
Note, however, that whereas Theorem\-~\ref{theo:Hartree} holds only for potentials of positive type (that is the Fourier transform of the potential $\hat v$ is non-negative), (\ref{high_density}) holds regardless of the sign of $\hat v$.
The asymptotic agreement at high densities therefore only holds for positive type potentials.
\bigskip

\indent
The proof of Theorem\-~\ref{theo:energy} proceeds as follows\-~\cite[Section 4]{CJL20}.
The high density formula\-~(\ref{high_density}) is easy to prove, and follows from the fact that $\int dx\ u(x)=\frac1\rho$.
To prove the low density expansion, we proceed in two steps: we first prove that the solution $u$ is close to the function $w$ defined by
\begin{equation}
  -\Delta w=(1-u)v
\end{equation}
and then we show that $w$ is close to the scattering solution $\varphi$, which solves
\begin{equation}
  -\Delta\varphi=(1-\varphi)v
  .
\end{equation}
To prove the first of these, we work in Fourier space, where we find that small $\rho$ corresponds to large momentum, so we can proceed by Taylor expansions.
The second follows from direct computation.
\bigskip

\begin{figure}
  \hfil\includegraphics[width=8cm]{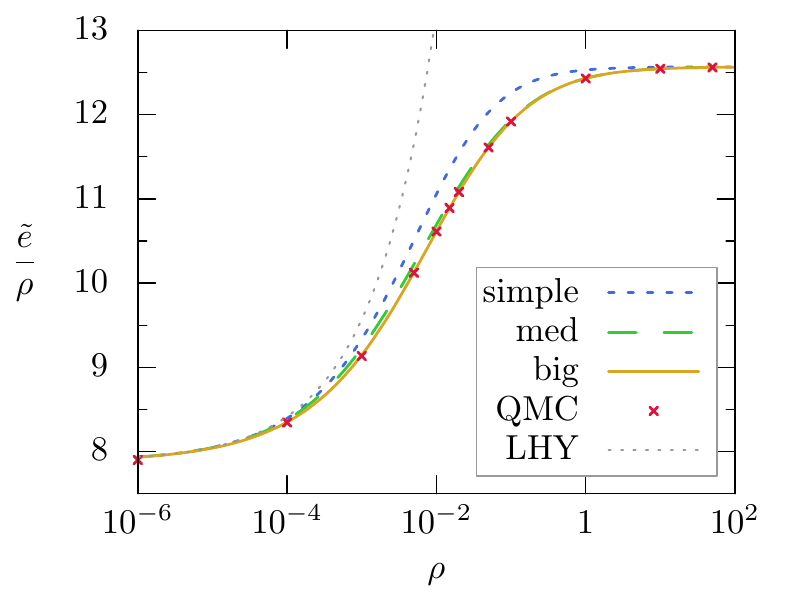}
  \caption{%
    \cite[Fig 1]{CHe21} The predictions of the energy per particle as a function of the density for the \eqformat{Simple Equation}, \eqformat{Medium Equation}, and \eqformat{Big Equation}, compared to a \eqformat{Quantum Monte Carlo} (QMC) simulation and the Lee-Huang-Yang (LHY) formula.
  }
  \label{fig:energy}
\end{figure}

\indent
Thus, the Simple Equation predicts the energy of the Bose gas at low and high densities (when the potential is of positive type).
A natural next question is to study the prediction at intermediate densities.
When $\rho$ is neither large nor small, it becomes much more difficult to evaluate the energy exactly.
Instead, we developed a software package, which we called {\tt simplesolv}\-~\cite{ss}, to compute the energy, and various other observables, numerically with high accuracy.
For a more detailed discussion of the accuracy of the numerics, see Appendix\-~\ref{app:simplesolv}.
For the graphs shown here, we chose the positive type potential $v(x)=e^{-|x|}$.
The result is reported in Figure\-~\ref{fig:energy}.
We computed the energy for the three equations in the Simplified Approach, and compared it to the prediction of a Quantum Monte Carlo simulation (QMC)\-~\cite{CHe21}.
While the Simple Equation agrees with the QMC at low and high densities, it only does somewhat well at intermediate densities (with a maximal relative error of approximately 5\%).
However, the Medium Equation fits the QMC data significantly better (with a maximal relative error of around 1\%), and the Big Equation provides a remarkably good fit at {\it all} densities (with a maximal relative error of approximately 0.1\%).
This indicates that the Simplified Approach, and the Big Equation in particular, are capturing some of the physics of the Bose gas at {\it all} densities, at least in the case of positive type potentials.
\bigskip

\subsection{Condensate fraction}
\indent
Whereas the Simplified Approach gives a natural prediction for the ground state energy, the condensate fraction must be computed in a more indirect way.
Indeed, the Simplified Approach computes $u$, see\-~(\ref{udef}) and\-~(\ref{g}), which is linear in $\psi$, whereas the condensate fraction is quadratic in $\psi$, see\-~(\ref{etadef}).
However, as was shown in\-~(\ref{eta_energy}), the condensate fraction can also be computed as the derivative of the ground state energy of a modified Hamiltonian.
The procedure to compute the condensate fraction in the Simplified Approach is to re-derive the equations of the Simplified Approach for the modified Hamiltonian, and then differentiate the prediction of the energy $\tilde f(\mu)$ that ensues:
\begin{equation}
  \tilde\eta=1+\partial_\mu \tilde f(\mu)|_{\mu=0}
  .
\end{equation}
We find that the effect of modifying the Hamiltonian on the Simplified Approach equations is to formally add $2\mu$ to $-\Delta$ in\-~(\ref{compleq})\-~\cite[(35)]{CHe21}:
\begin{equation}
  -\Delta u(x)
  +2\mu u(x)
  =
  (1-u(x))\left(v(x)-2\rho K(x)+\rho^2 L(x)\right)
\end{equation}
and consequently, for the Simple Equation,
\begin{equation}
  -\Delta u(x)+2\mu u(x)=(1-u(x))v(x)-4\tilde e u(x)+2\tilde e\rho u\ast u(x)
  .
  \label{simpleq_mu}
\end{equation}
\bigskip

\indent
Whereas the condensate fraction has not been evaluated exactly for the Bose gas, even at low densities, Bogolyubov theory predicts an asymptotic formula which we stated in Conjecture\-~\ref{conj:condensate}.
For the Simple Equation, this formula can be shown to hold.
\bigskip

\theoname{Theorem}{\cite[Theorem 1.6]{CJL21}}
  For the Simple Equation, if $(1+|x|^4)v(x)\in L_1(\mathbb R^3)\cap L_2(\mathbb R^3)$ and $v\geqslant 0$, as $\rho\to0$,
  \begin{equation}
    \tilde\eta\sim\frac{8\sqrt{\rho a^3}}{3\sqrt\pi}
  \end{equation}
  where $a$ is the scattering length of the potential\-~\cite[Appendix C]{LSe05}.
\endtheo
\bigskip

To prove this theorem, we compute $\partial_\mu u$ by differentiating\-~(\ref{simpleq_mu}) with respect to $\mu$, and find\-~\cite[Theorem\-~1.6, Section\-~5]{CJL21}
\begin{equation}
  \tilde\eta=\frac{\rho\int dx\ v(x)\mathfrak K_{\tilde e}u(x)}{1-\rho\int dx\ v(x)\mathfrak K_{\tilde e}(2u-\rho u\ast u)(x)}
\end{equation}
where $\mathfrak K_{\tilde e}$ is defined in\-~(\ref{frakK}).
We then switch to Fourier space, where, as for the energy, small $\rho$ corresponds to large momentum, and compute these integrals.
\bigskip

\begin{figure}
  \hfil\includegraphics[width=8cm]{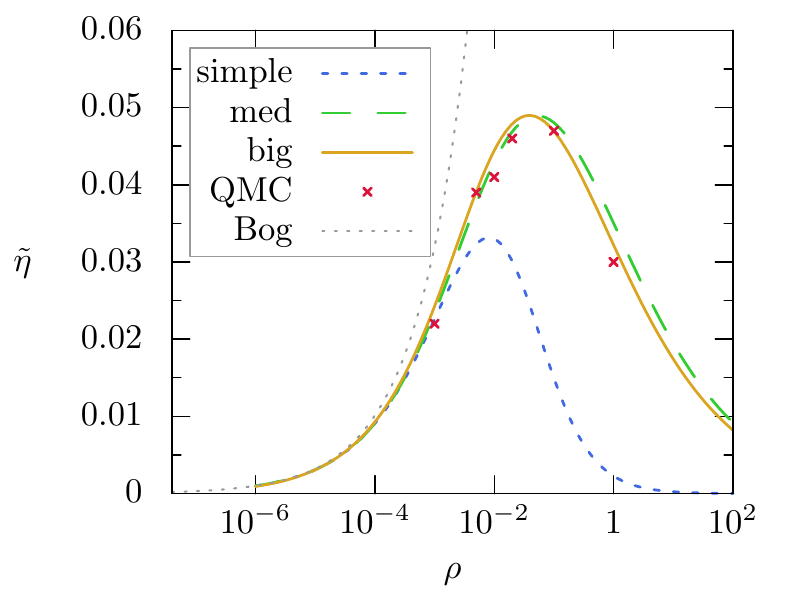}
  \caption{%
    \cite[Fig 3]{CHe21} The predictions of the condensate fraction as a function of the density for the \eqformat{Simple Equation}, \eqformat{Medium Equation}, and \eqformat{Big Equation}, compared to a \eqformat{Quantum Monte Carlo} (QMC) simulation and the Bogolyubov prediction (Bog).
  }
  \label{fig:condensate_fraction}
\end{figure}

\indent
The Simple Equation yields the same prediction as Bogolyubov theory at low density.
As we did for the ground state energy, we will now turn our attention to intermediate densities, though, here again, we will have to limit ourselves to a numerical analysis using {\tt simplesolv}\-~\cite{ss}.
(Note that, in {\tt simplesolv}, the derivative with respect to $\mu$ is carried out formally, and not numerically as a finite difference, and the condensate fraction is expected to be computed just as accurately as the energy.)
For this graph, we take a slightly smaller potential: $v(x)=\frac12e^{-|x|}$ (as the potential gets larger, the agreement between the Simplified Approach and the Bose gas gets worse and worse, so we chose a smaller potential to highlight the difference in the accuracies of the various equations in the Simplified Approach).
The result is reported in Figure\-~\ref{fig:condensate_fraction}.
We find that for all of the equations of the Simplified Approach, the predictions for the condensate fraction agree with Bogolyubov theory at low density, and return to full condensation ($\tilde\eta=0$, recall that $\eta$ is the {\it un}condensed fraction) at high density.
At intermediate density, the Simple Equation does not do very well, but the Medium and Big Equations are rather accurate, though not nearly as accurate as for the energy.
We note that the uncondensed fraction goes through a maximum, which is reproduced fairly accurately by the Medium and Big Equations.
The physical significance of this maximum is yet to be investigated.
As we will see in the next subsection, looking at the two-point correlation function near the maximal density reveals some non-trivial physical behavior in certain cases.
\bigskip

\subsection{Two-point correlation function}
\indent
Similarly to the condensate fraction, the two-point correlation function cannot be evaluated directly from $u$, but instead is computed by differentiating the energy: following the procedure in\-~(\ref{Cderiv}), it is natural to define the prediction of the Simplified Approach for the two-point correlation function as
\begin{equation}
  \tilde C_2(z)=2\rho\frac{\delta\tilde e}{\delta v(z)}
  .
  \label{C2prediction}
\end{equation}
\bigskip

\indent
As was the case for the condensate fraction, there are no exact estimates for the two-point correlation function of the Bose gas, but the large distance behavior has been predicted to be $|x|^{-4}$ (see Conjecture\-~\ref{conj:2pt}).
We can prove that this is the case for the prediction of the Simple Equation.
\bigskip

\theo{Theorem}\label{theo:2pt}
  Under the assumptions of Theorem\-~\ref{theo:existence}, and if $(1+|x|)^6v(x)\in L_1(\mathbb R^3)$,
  \begin{equation}
    \lim_{|x|\to\infty}|x|^4\left|\frac{\tilde C_2}{\rho^2}-1-r(x)\right|<\infty
  \end{equation}
  where $|x|^4r\in L_2(\mathbb R^3)\cap L_\infty(\mathbb R^3)$.
\endtheo
\bigskip

This result has not been published until now, but its proof can be found in Appendix\-~\ref{app:decay_2pt}.
It is based on a formula for $\tilde C_2$ from\-~\cite[(45)]{CHe21}:
\begin{equation}
  \tilde C_2(x):=
  \rho^2(1-u(x))
  +\rho^2\frac{\mathfrak K_{\tilde e}v(x)(1-u(x))-2\rho u\ast \mathfrak K_{\tilde e}v(x)+\rho^2u\ast u\ast \mathfrak K_{\tilde e}v(x)}{1-\rho\int dx\ v(x)\mathfrak K_{\tilde e}(2u(x)-\rho u\ast u(x))}
\end{equation}
where $\mathfrak K_{\tilde e}$ is defined in\-~(\ref{frakK}), which one obtains by taking the functional derivative with respect to $v$ in\-~(\ref{simpleq}).
\bigskip

\begin{figure}
  \hfil\includegraphics[width=8cm]{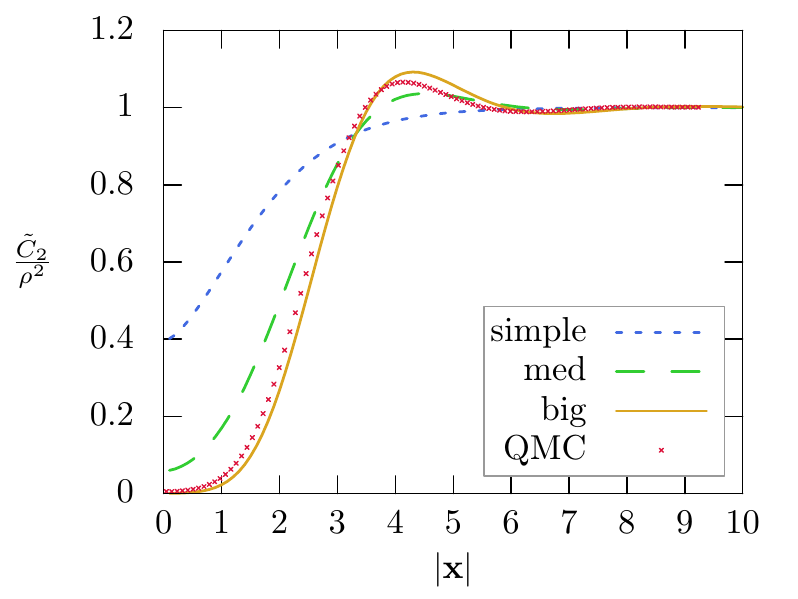}
  \caption{%
    \cite[Figure\-~5]{CHe21}
    The predictions of the two-point correlation function as a function of $x$ at $\rho=0.02$ for $v(x)=16e^{-|x|}$ for the \eqformat{Simple Equation}, \eqformat{Medium Equation}, and \eqformat{Big Equation}, compared to a \eqformat{Quantum Monte Carlo} (QMC) simulation.
  }
  \label{fig:2pt}
\end{figure}

\indent
Using\-~(\ref{C2prediction}), it is rather easy (and quick) to compute the prediction of the two-point correlation function for the equations in the Simplified Approach numerically.
In doing so, we found an intriguing physical phenomenon at intermediate densities.
The phenomenon was observed for the potential $v(x)=16e^{-|x|}$ (and seems to be absent for $e^{-|x|}$).
The result is shown in Figure\-~\ref{fig:2pt}.
We find that for the Medium and Big Equations, there is a higher probability of finding two particles separated by a certain length than any other.
This indicates a phase resembling a liquid, in which there is no long range order, but particles are likely to find themselves separated by similar distances.
We have verified that this phenomenon also occurs for the Monte Carlo simulation of the Bose gas.
It is notably absent from the prediction of the Simple Equation though, which provides further evidence that the Big and Medium Equations reproduce the physics of the Bose gas more accurately than the Simple Equation.
The phenomenon also disappears for smaller as well as larger densities, which is consistent with the presence of a phase transition.
While our analysis of this phase is still in its early stages, this phenomenon is is rather intriguing, and shows how interesting the intermediate density regime might be.
\bigskip

\section{Conclusion and open problems}\label{sec:conclusion}
\subsection{Summary}
\indent
The Bose gas is a system that is simple to define, and yet is an accurate model for some real-world physical systems.
It has a rich phenomenology, and has proven to be quite difficult to study, at least in the presence of interactions.
We have discussed a family of effective equations for the Bose gas which are much simpler to study, both analytically (for the Simple Equation) and numerically, and reproduce some of the intriguing features of the Bose gas.
What is unique about these effective equations is that they are accurate {\it both} at low {\it and} at high densities, and have even given us insight on the behavior of the system at intermediate densities.
\bigskip

\indent
Among these equations, the only one we can study analytically is the Simple Equation.
This equation yields a prediction for the ground state energy of the Bose gas which reproduces all the known results about the ground state energy of the Bose gas, both at low and high densities (for potentials of positive type).
Furthermore, we have proved that it reproduces the Bogolyubov prediction for the condensate fraction, as well as the rate of decay of two-point correlation functions.
On the other hand, the approach used to derive the Simple Equation is quite different from Bogolyubov theory, so there is hope that it will give us a new approach to proving properties for the Bose gas, such as Bose-Einstein condensation.
\bigskip

\indent
Deriving the Simple Equation goes through several uncontrolled approximations, and we have introduced two other equations which are still simple enough to be solved numerically efficiently, but require fewer approximations, and so reproduce the physics of the Bose gas more accurately.
We have found that they both (to varying degrees) give good qualitative and quantitative agreement for potentials of positive type, in the {\it entire} range of densities.
In particular, the Medium and Big Equations provide us with useful tools to study the Bose gas at intermediate densities, in a regime that has not been studied in the past.
We have found promising preliminary results about the physics of the Bose gas at intermediate densities, which suggests the presence of a liquid-type phase.
\bigskip

\indent
The assumptions on the potential required for the Simplified Approach to provide accurate results are that $v(x)\geqslant 0$, $v\in L_1(\mathbb R^3)\cap L_{\frac32+\epsilon}(\mathbb R^3)$ (some results also require stronger decay properties, see above), and that the potential be of positive type (its Fourier transform should be non-negative).
The latter assumption is rather restrictive, and excludes a number of physical systems such as Helium IV gases.
On the other hand, from a mathematical point of view, it is easy to construct such potentials.
As stated here, our results require the potential not to have a hard core, however, most of the results in the present paper can be extended easily to the hard core case.

\subsection{Open problems}
\indent
There are still a number of interesting unanswered questions about the Simplified Approach to the Bose gas.
One is the uniqueness of the solution of the Simple Equation.
As was explained in Section\-~\ref{sec:uniqueness}, we currently only have a weak notion of uniqueness, in which a density $\rho$ could yield several different values of the energy.
To prove the uniqueness, it would suffice to prove that the energy is a monotone increasing function of the density, which is true for the repulsive Bose gas.
We have proved the monotonicity of the energy only for small and large densities\-~\cite[Theorem\-~1.3]{CJL21}.
A related question is to prove that $\rho e(\rho)$ is a convex function of $\rho$.
Physically, this is equivalent to the compressibility of the gas being positive (which is certainly the case).
We have proved this convexity only for small density\-~\cite[Theorem\-~1.5]{CJL21}.
\bigskip

\indent
All the analytical results we have managed to prove so far concern the Simple Equation, but as we saw, the Medium and Big Equations reproduce the physics of the Bose gas much more accurately.
An interesting open problem is to prove the existence of solutions for these equations.
Numerically, we construct the solutions using the Newton algorithm.
It would suffice to prove that this algorithm converges, and we have found in practice that it does so for a wide variety of initializations.
\bigskip

\indent
Another direction to investigate is whether the Simplified Approach can yield results for the Bose gas.
For instance, we have observed numerically that the Simple Equation predicts an energy which is larger than the energy of the Bose gas.
If such a statement could be proved, it would provide us with an upper bound on the energy that would hold at all densities.
It would also give us the Lee-Huang-Yang formula\-~(\ref{lhy}) as an upper bound at low densities, and since the Simple Equation can be studied for hard core potentials, it would fill in a gap in our current knowledge on the ground state energy at low densities.
The difficulty here is that the Simplified Approach does not provide an Ansatz for the ground state wavefunction.
Such an Ansatz could be obtained from the solution $u$ using a Bijl-Dingle-Jastrow function\-~\cite{Bi40,Di49,Ja55} (sometimes referred to simply as a Jastrow function), but evaluating the energy of such a wavefunction is not an easy task.
\bigskip

\indent
An important open problem is to estimate the error made by the approximations leading to the Simplified Approach.
This would allows us to estimate how far the predictions of the Simplified Approach are from the Bose gas.
Another intriguing open question is whether the Simplified Approach can be extended to excited states, or to a thermal state.
However, the derivation of the Simplified Approach only specializes to the ground state through the fact that the wavefunction is non-negative, which allows for the probabilistic interpretation of the correlation functions $g_N^{(n)}$ to hold.
At the moment, it is not clear how to adapt this to excited states.

\appendix

\section{Some comments on numerical computations}\label{app:simplesolv}
\indent
To carry out the numerical computations, we wrote a software package called {\tt simplesolv}\-~\cite{ss}, which is available for download from\par
\hfil{\tt\color{blue}\href{http://ian.jauslin.org/software/simplesolv}{http://ian.jauslin.org/software/simplesolv}}\par
and is released under the Apache 2.0 open source license.
It is written in the {\tt julia} programming language, and requires the {\tt julia} interpreter to run.
The documentation that is bundled with the software contains information on how to use the software, as well as detailed descriptions of the computations involved.
\bigskip

\indent
{\tt simplesolv} can compute the solution to any of the equations in the Simplified Approach (as well as any interpolation between the Complete Equation and the Simple Equation.
It can compute various observables: the ground state energy per particle, the condensate fraction, the two-point correlation function, and the momentum distribution, as well as the solution $u(x)$ and its Fourier transform $\hat u(k)$.
Seven different potentials are built in, and it is rather simple to program custom potentials.
In addition, {\tt simplesolv} can compute solutions of the Simple Equation with a hard core potential.
\bigskip

\indent
The computations are carried out in Fourier space, where there are fewer convolutions.
Momentum space is compactified using the transformation $|k|\mapsto\frac{1-|k|}{1+|k|}$.
Integrals are computed using Gauss quadratures, which consist in a replacing integrals by sums, in which the integrands are evaluated at carefully chosen points $|k_i|$.
The error of Gauss quadratures decays exponentially with the number of sample points (provided the integrand is analytic).
In the case of the Simple and Medium Equations, we represent the solution $\hat u$ as a vector whose components are $\hat u(k_i)$, where the $k_i$ are the points defined by the Gauss quadrature algorithm.
Due to the fact that these equations do not involve convolutions in Fourier space, the solution $\hat u$ only ever needs to be evaluated at $k_i$, and so representing $\hat u$ by this finite-dimensional vector works well.
In the case of the Big Equation, which involves convolutions in Fourier space, we need an interpolation scheme to evaluate $\hat u$ away from $k_i$.
We chose the Chebyshev polynomial approximation, whose error in $L_\infty$ norm decays exponentially with the order of the polynomials (in the case of analytic functions).
In order to avoid boundary effects propagating to all momenta, we split momentum space into intervals (called ``splines''), and expand into Chebyshev polynomials inside each interval.
The Complete Equation is treated similarly, although computation times increase dramatically.
\bigskip

\indent
Having specified the solution $\hat u$ by a finite-dimensional vector, we solve the equations using the Newton algorithm, which has a super-exponential rate of convergence.
The algorithm is run until the Newton step becomes smaller than a specified tolerance.
For the Simple and Medium Equations, the Newton algorithm is initialized with the scattering solution, whereas for all other equations, it is initialized with the solution of the Medium Equation.
\bigskip

\indent
There is a significant difference in the run-time for the different equations.
The Simple and Medium Equations are significantly faster than the Big Equation, which itself is significantly faster than the Complete Equation.
As we saw in the discussion above, the Medium Equation is rather accurate, and it is one of the quickest and easiest to solve numerically.
\bigskip

\indent
For readers interested in reproducing the numerical results presented here and in\-~\cite{CHe21}, the values of all parameters are available, bundled with the preprints, at the following websites: for\-~\cite{CHe21}:\par\penalty10000
\hfil{\tt\color{blue}\href{http://ian.jauslin.org/publications/20chjl}{http://ian.jauslin.org/publications/20chjl}}\par
and for the present paper:\par\penalty10000
\hfil{\tt\color{blue}\href{http://ian.jauslin.org/publications/22j}{http://ian.jauslin.org/publications/22j}}\par
In each case, the relevant information can be found in the file {\tt ./figs/*.fig/Makefile} where the commands that were run to obtain graphs are written out.

\section{Decay of the two-point correlation function}\label{app:decay_2pt}
\indent
In this appendix, we prove Theorem\-~\ref{theo:2pt}.
\medskip

\indent
  Let
  \begin{equation}
    \sigma:=
    \frac1{1-\rho\int dx\ v(x)\mathfrak K_{\tilde e}(2u(x)-\rho u\ast u(x))}
  \end{equation}
  in terms of which
  \begin{equation}
    \frac{\tilde C_2}{\rho^2}-1=-u(x)-\sigma u(x)\mathfrak K_{\tilde e}v(x)+\sigma\Phi(x)
  \end{equation}
  with
  \begin{equation}
    \Phi(x):=\mathfrak K_{\tilde e}v(x)-2\rho u\ast \mathfrak K_{\tilde e}v(x)+\rho^2 u\ast u\ast \mathfrak K_{\tilde e}v(x)
    =\mathfrak K_{\tilde e}v\ast(\delta-\rho u)\ast(\delta-\rho u)(x)
  \end{equation}
  where $\delta(x)$ is the Dirac delta function.
  Note that $|x|^4u\in L_\infty(\mathbb R^3)$ (see Theorem\-~\ref{theo:decay}).
  Furthermore, $\mathfrak K_{\tilde e}v\in L_2(\mathbb R^3)\cap L_\infty(\mathbb R^3)$ (see\-~\cite[Lemma\-~1.10]{CJL21}), so
  \begin{equation}
    |x|^4\left(\frac{\tilde C_2(x)}{\rho^2}-(1-u(x))-\sigma\Phi(x)\right)=-|x|^4\sigma\mathfrak K_{\tilde e}v(x)=:|x|^4r(x)\in L_2(\mathbb R^3)\cap L_\infty(\mathbb R^3)
    .
    \label{term1}
  \end{equation}
  We now turn to the asymptotics of $\Phi$.
  \bigskip

  \point
  First note that, by the resolvant identity,
  \begin{equation}
    \mathfrak K_{\tilde e}v=\mathfrak Y_{\tilde e}(v(1-\mathfrak K_{\tilde e}v))
    ,\quad
    \mathfrak Y_{\tilde e}:=
    (-\Delta+4\tilde e(1-\rho u\ast))^{-1}
    .
  \end{equation}
  Let
  \begin{equation}
    w:=v(1-\mathfrak K_{\tilde e}v)
  \end{equation}
  which, by\-~\cite[Lemma\-~1.10]{CJL21}, satisfies $0\leqslant w\leqslant 1$ and $w\in L_1(\mathbb R^3)$.
  In terms of $w$,
  \begin{equation}
    \Phi(x)
    =\mathfrak Y_{\tilde e}w\ast(\delta-\rho u)\ast(\delta-\rho u)(x)
    =w\ast(\delta-\rho u)\ast\mathfrak Y_{\tilde e}(\delta-\rho u)(x)
    .
  \end{equation}
  We will now compute the decay of
  \begin{equation}
    \Psi(x):=\mathfrak Y_{\tilde e}(\delta-\rho u)
  \end{equation}
  and show that
  \begin{equation}
    |\Psi(x)|\leqslant\frac{(\mathrm{const}.)}{|x|^6}
    .
  \end{equation}
  In Fourier space,
  \begin{equation}
    4\tilde e\hat\Psi=\frac{1-\rho\hat u}{\kappa^2+(1-\rho\hat u)}
    =\frac1{1+\frac{\kappa^2}{1-\rho\hat u}}
    ,\quad
    \kappa:=\frac{|k|}{2\sqrt{\tilde e}}
    .
  \end{equation}
  We bound
  \begin{equation}
    ||x|^6\Psi|
    \leqslant
    \|\Delta^3_k\hat\Psi\|_1
    =
    \frac{2\pi}{\sqrt{\tilde e}} \int d\kappa\ \kappa^2\left|\partial_\kappa^6\hat\Psi+\frac6\kappa\partial_\kappa^5\hat\Psi\right|
    .
  \end{equation}
  \bigskip

  \point
  We bound $\partial_\kappa^n\hat\Psi$.
  \medskip

  \subpoint
  We first compute $\partial_\kappa^n\hat\Psi$ for $\kappa\ll 1$.
  We have\-~\cite[(4.25)]{CJL20}
  \begin{equation}
    \rho\hat u=\kappa^2+1-\sqrt{(\kappa^2+1)^2-\frac\rho{2\tilde e}\hat S}
    ,\quad
    S(x):=(1-u(x))v(x)
  \end{equation}
  and since $(1+|x|)^6v\in L_1(\mathbb R^3)$,
  \begin{equation}
    \frac\rho{2\tilde e}\hat S(\kappa)=1-\beta_2\kappa^2+\beta_4\kappa^4+\beta_6\kappa^6+O(\kappa^8)
  \end{equation}
  with $\beta_2>0$.
  By a direct computation, we find that
  \begin{equation}
    4\tilde e\hat\Psi(0)=1
    ,\quad
    4\tilde e\partial_\kappa^6\hat\Psi|_{\kappa=0}=0
    ,\quad
    4\tilde e\partial_\kappa^5\hat\Psi|_{\kappa=0}=
    \frac{3\beta_4^2-6\beta_4+3+(4\beta_2+8)\beta_6}{8(2+\beta_2)^{\frac52}}
    \label{Psi0}
  \end{equation}
  so, for $5\leqslant n\leqslant 6$
  \begin{equation}
    \partial_\kappa^n\hat\Psi=O(1)
  \end{equation}
  (from here on out, all constants may depend on $\tilde e$).
  \bigskip

  \subpoint
  We now turn to $\kappa\gg1$.
  First note that, since $(1+|x|)^6v\in L_1(\mathbb R^3)$, for $0\leqslant n\leqslant 4$,
  \begin{equation}
    \frac\rho{2\tilde e}|\partial_\kappa^nS|=O(1)
    .
  \end{equation}
  By a direct evaluation we find that
  \begin{equation}
    4\tilde e\hat\Psi=\frac1{\kappa^2}+O(\kappa^{-4})
    ,\quad
    4\tilde e\partial_\kappa^5\hat\Psi
    =-\frac{\frac\rho{2\tilde e}\partial_\kappa^5\hat S}{2\kappa^4}+O(\kappa^{-5})
    ,\quad
    4\tilde e\partial_\kappa^6\hat\Psi
    =-\frac{\frac\rho{2\tilde e}\partial_\kappa^6\hat S}{2\kappa^4}+O(\kappa^{-5})
    \label{Psiinf}
  \end{equation}
  so
  \begin{equation}
    \left|\partial_\kappa^6\hat\Psi+\frac6\kappa\partial_\kappa^5\hat\Psi\right|
    =O(\kappa^{-4})
    .
  \end{equation}
  \bigskip

  \subpoint
  Thus $|x|^6\Psi\in L_\infty(\mathbb R^3)$, and so
  \begin{equation}
    \Psi\leqslant\frac{(\mathrm{const}.)}{|x|^6}
    .
  \end{equation}
  \bigskip

  \point
  Furthermore, if $(1+|x|)^6v\in L_1(\mathbb R^3)$, then
  \begin{equation}
    |w(x)|\leqslant\frac{(\mathrm{const}.)}{|x|^6}
    .
  \end{equation}
  Now, given two functions $f,g$ such that $f\sim\alpha|x|^{-n}$ and $g\sim\beta|x|^{-m}$ with $m>n$, $f\ast g\sim\alpha|x|^{-n}\int dx\ g(x)$, so
  \begin{equation}
    \Phi\equiv w\ast(\delta-\rho u)\ast\Psi\sim_{|x|\to\infty}
    -\rho u\int dx\ w\ast\Psi(x)
    .
  \end{equation}
  Furthermore, if $v\in L_{\frac32+\epsilon}(\mathbb R^3)$, then $w\in L_{\frac32+\epsilon}(\mathbb R^3)$, and by\-~(\ref{Psi0}) and\-~(\ref{Psiinf}), $\hat\Psi\in L_{\frac{3+2\epsilon}{1+2\epsilon}}(\mathbb R^3)$, so
  \begin{equation}
    \left\|\int dx\ w\ast\Psi(x)\right\|
    \leqslant
    \|w\|_{\frac32+\epsilon}\|\Psi\|_{\frac{3+2\epsilon}{1+2\epsilon}}
    .
  \end{equation}
  Thus,
  \begin{equation}
    \Phi
    \sim (\mathrm{const}.)u
    .
  \end{equation}
  We conclude the proof using\-~(\ref{term1}) and Theorem\-~\ref{theo:decay}.
\qed

\vfill
\eject

\end{document}